\documentclass[draft]{agujournal}
\journalname{JGR-Space Physics}
\usepackage{float}
\begin{document}
\justify
\title{Multiple satellite analysis of the Earth's thermosphere and interplanetary magnetic field variations due to ICME/CIR events during 2003--2015}

\authors{S. Krauss\affil{1,*}, M. Temmer\affil{2}, S. Vennerstrom\affil{3}}
\affiliation{1}{Space Research Institute, Austrian Academy of Sciences}
\affiliation{2}{Institute of Physics, University of Graz}
\affiliation{3}{Institute of Astrophysics and Atmospheric Physics, Technical University of Denmark}
\affiliation{*}{now at the Institute of Geodesy, Graz University of Technology}
\correspondingauthor{Sandro Krauss}{sandro.krauss@tugraz.at}

\begin{keypoints}
 \item = This study depicts the response of the thermosphere to coronal mass ejections and corotating interaction regions at two different altitudes.
 \item = In total, the analysis comprises nearly 400 solar events, which took place from 2003 to 2015.
 \item = Resulting satellite orbit decays are evaluated for the different orbital altitudes and solar events with varying intensity.
\end{keypoints}

\begin{abstract}
We present a refined statistical analysis based on interplanetary coronal mass ejections as well as co-rotating interaction regions for the time period 2003--2015 to estimate the impact of different solar wind types on the geomagnetic activity and the neutral density in the Earth's thermosphere.  For the time-based delimitation of the events, we rely on the catalog maintained by Richardson and Cane and the CIR-lists provided by S. Vennerstrom and~\citet{Jian2011}. These archives are based on in-situ measurements from the ACE and/or the Wind spacecraft.
On this basis, we thoroughly investigated 196 Earth-directed ICME and 195 CIR events. To verify the impact on the Earth’s thermosphere we determined neutral mass densities by using accelerometer measurements collected by the low-Earth orbiting satellites GRACE and CHAMP. Subsequently, the atmospheric densities are related to characteristic ICME parameters. In this process a new calibration method has been examined. Since increased solar activity may lead to a decrease of the satellites orbital altitude we additionally assessed the orbital decay for each of the events and satellites.\\
The influence of CIR events is in the same range of magnitude as the majority of the ICMEs (186 out of 196). Even though, the extended investigation period between 2011 and 2015 has a lack of extreme solar events the combined analysis reveals comparable correlation coefficients between the neutral densities and the various ICME and geomagnetic parameters (mostly $> 0.85$). The evaluation of orbit decay rates at different altitudes revealed a high dependency on the satellite actual altitude.
\end{abstract}

\section{Introduction}
Accompanying with the rapid technological progress in the last decades the topic "space weather" is becoming extremely important for life in our society. Space weather disturbances in terms of geomagnetic storms, aurorae or geomagnetically induced currents (GIC) can affect space-borne and ground based technological systems. The strong effect of turbulent magnetic field from interplanetary coronal mass ejection (ICMEs) on the Earth's magnetosphere, driving strong geomagnetic storms and GICs \citep[see][]{Huttunen2008,Guo2011,Lugaz2016}. In a previous study~\citep{Krauss2015} the thermospheric and geomagnetic response to 104 ICMEs observed by the Advanced Composition Explorer (ACE;~\cite{Stone1998}) and Gravity Recovery and Climate Experiment (GRACE;~\citet{Tapley2004}) satellites during the period July 2003 to August 2010 has been analyzed. It was found that the strength of the $B_{\rm z}$ component, either in the shock-sheath or in the magnetic structure~\citep{Burlaga1982} of the ICME, is highly correlated with the neutral density enhancement. An increase in thermospheric neutral densities affects Earth-orbiting satellites in such a way that the drag force on the spacecraft is enhanced, which may lead to severe consequences for the satellite's altitude. For example, during the ICME event on October 29, 2003 the orbit decay for the International Space Station (ISS) for one day increased by a factor of 4~\citep{Bean2007}. Therefore, enhanced statistics are needed in order to better estimate the response of the thermospheric density to impacting solar phenomena. In that respect, not only ICMEs but also co-rotating interaction regions (CIRs) are of interest~\citep{Chen2014}. The interaction between slow and fast solar wind streams, emanating from coronal holes, results in compression regions. Those regions are related to a very turbulent and strong magnetic field and usually last longer compared to the shock-sheath region of a ICME. However, the negative peak in the $B_{\rm z}$ component of the magnetic field is much larger. Within a CIR the plasma pressure peaks close to the stream interface region, which is accompanied by a large shear flow~\citep{Gosling1999}. Especially during low solar activity phases of the solar cycle, CIR induced storms occur very frequently and quasi-periodically~\citep{Temmer2007,Lei2008}. Even though, the compression region of CIRs might be less energetic compared to those of ICMEs, it lasts longer and it is a recurrent phenomenon. Following~\citet{Turner2009}, magnetic storms driven by CIRs seem to deposit more energy in the ionosphere and ring current and thus appear to be more geoefficient than ICMEs.

In this paper, we rely on an extended dataset of thermospheric density measurements compared to our predecessor study. At that time, the investigation period was rather limited due to missing calibration parameters for the GRACE accelerometer data beyond 2010. However, through the development of a new calibration method~\citep{Klinger2016}, the analysis period for the current study could be nearly doubled (2003--2015).
\begin{figure*}[htb]
 \begin{center}
  \includegraphics[width=8cm]{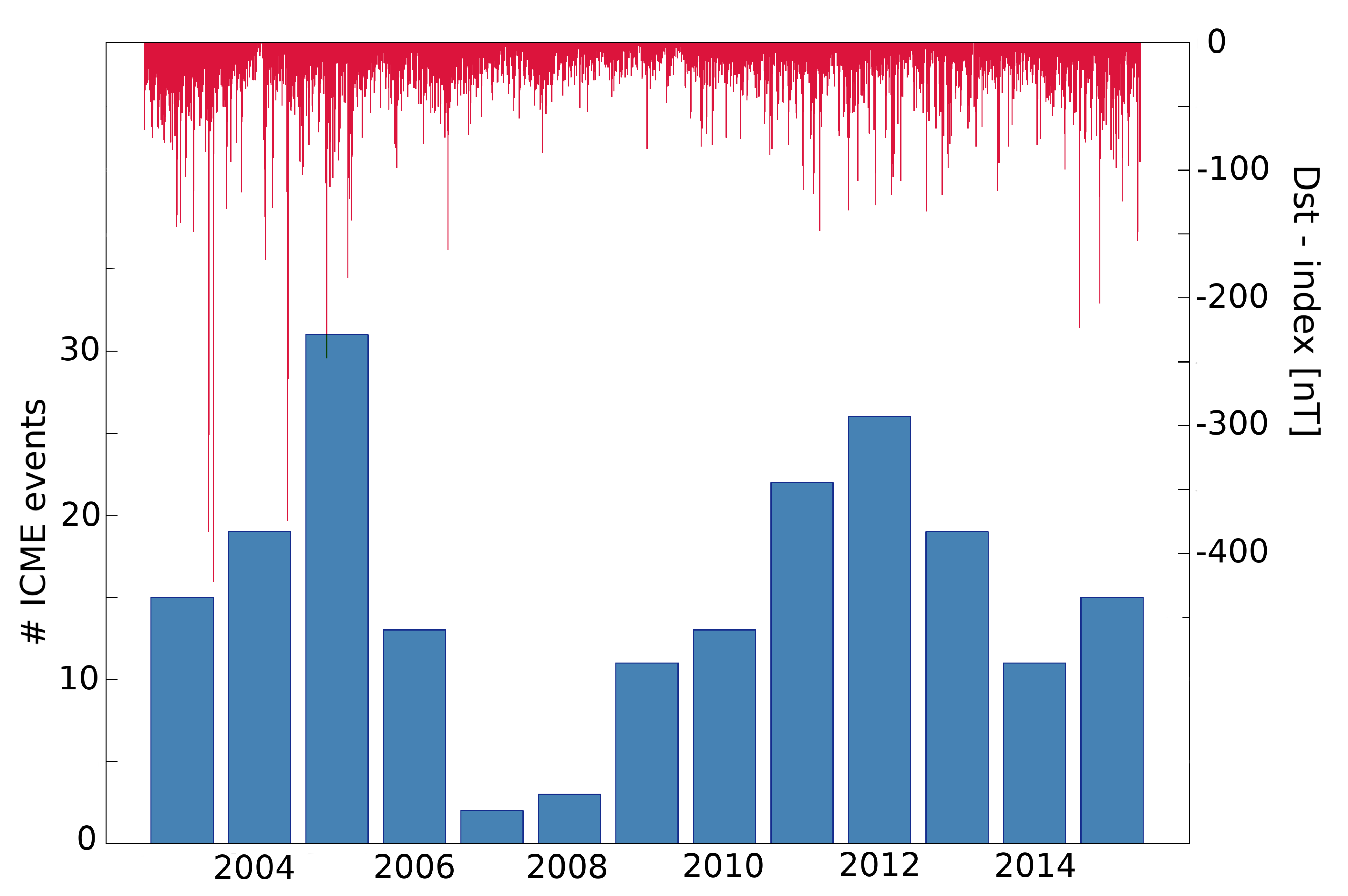}
 \end{center}
\caption{Temporal distribution of the analyzed ICME events (blue bars) between 2003 and 2015. The red bars indicate the level of geomagnetic activity in terms of the geomagnetic disturbance storm time index (Dst)~\citep{Sugiura1964,SugiuraKamei1991}.}
 \label{fig:1}
\end{figure*}
Against this background Figure~\ref{fig:1} illustrates an overview of all ICME events analyzed in this study at annual intervals. Even though, very strong geomagnetic storms during solar cycle 24 were missing, due to the lower solar activity level~\citep{Guo2011}, Figure~\ref{fig:1} clearly reveals that, through the temporal extension, nearly 100 additional ICME events could be included in the analysis. In total, the present study comprises measurements of about 400 solar events including the complementary data set of CIRs.

\section{Data and Analysis}
For the current study we analyzed geomagnetic storms induced by 196 ICME and 195 CIR events during the 13-years time period from 2003 to 2015. The basis for the selection of the ICME events forms the catalog of near-Earth interplanetary ICMEs maintained by~\citet{Richardson2010} (R\&C). Arrival times for CIR events are provided through reference lists by S. Vennerstrom and \citet{Jian2011}. Technically speaking, these lists cover stream interaction regions (SIR); the only difference between a SIR and a CIR is that the latter recurs for two or more solar rotations. The effects of these solar storms on the near-Earth environment are thoroughly analyzed by using in-situ magnetic field measurements performed by the ACE spacecraft and in addition by accelerometer measurements on board the low Earth orbiting satellites GRACE and Challenging Mini Payload Satellite (CHAMP;~\citet{Reigber2002}).
\subsection{Magnetic field measurements}
\begin{figure*}[htb]
 \begin{center}
  \includegraphics[width=8cm]{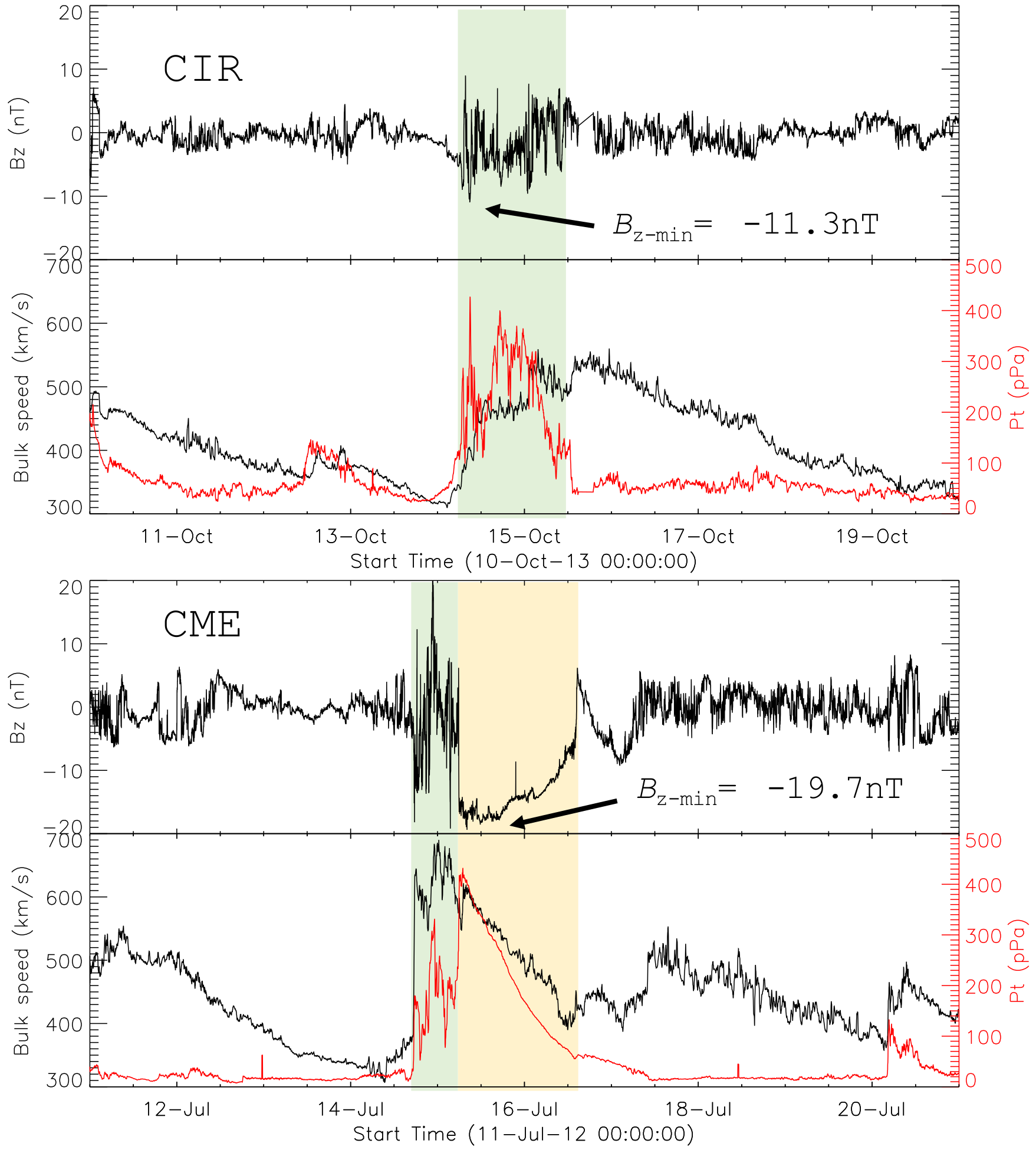}
 \end{center}
\caption{CIR (top) and ICME (bottom) magnetic field observation, total perpendicular pressure (sum of magnetic pressure and perpendicular plasma thermal pressure) and bulk speed measured by ACE. Green shaded area marks the compression region. Yellow shaded area marks the magnetic structure.}
 \label{fig:2}
\end{figure*}

In general, the evaluation of the ICME kinematics is similar to those described in~\citet{Krauss2015}. We process in-situ measurements (level 2 data) by the ACE spacecraft, which is located at the Lagrange point L1, in terms of solar wind flow speed (proton bulk speed) and the magnetic field component $B_{\rm z}$ (in Geocentric Solar Ecliptic, GSE, coordinates) with a time resolution of four minutes (OMNI database;~\citet{King2004}). The presumed starting times of the disturbance are taken from the R\&C catalog and act as a starting point for a 36h time window, from which the negative $B_{z}$ peak was extracted. The analysis of the 195 CIR events was performed by applying the same methods. In contrast to ICMEs and their rather smooth long-term negative $B_{\rm z}$ component in the magnetic structure, geomagnetic storms induced by CIRs are characterized by lower energy input but show long-persistent variations of the $B_{\rm z}$ component in the compression region. To illustrate this behavior Figure~\ref{fig:2} shows the $B_z$ and speed component of the solar wind, as well as the total perpendicular pressure ($P_t$) for a ICME and CIR event.

\subsection{Thermospheric neutral density}
\begin{figure*}[htb]
 \begin{center}
  \includegraphics[width=8cm]{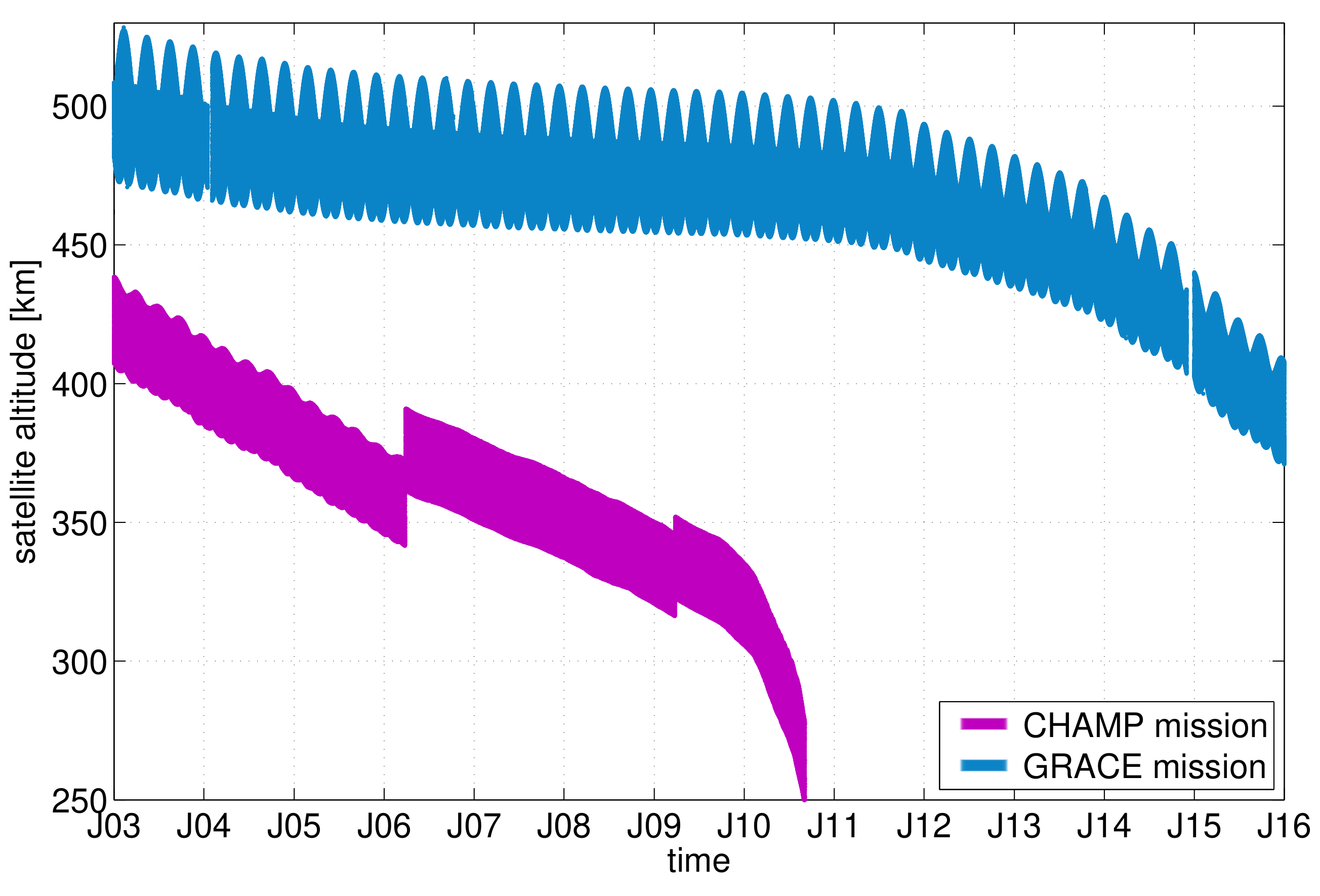}
 \end{center}
\caption{Course of the satellite altitudes of the CHAMP and GRACE spacecraft during the period of investigation (2003--2015).}
 \label{fig:3}
\end{figure*}

As it is generally known, the Earth's atmosphere density decreases exponentially with increasing height. The further away from the surface, the less weight is in the air column above. To examine the response of the Earth's thermosphere at different satellite altitudes we calculated neutral densities using accelerometer measurements from two satellite missions, namely CHAMP and GRACE. The former spacecraft represents the first dedicated Earth gravity field mission and was launched in July 2000 in a near-circular and near-polar orbit (eccentricity $\approx~0.004$, inclination $\approx 87^{\circ}$). During the ten-years mission duration the satellite altitude decreased steadily from the initially 454\,km to below 300\,km.
The second mission which we examined in our analysis is the U.S.-German project GRACE, which was launched in 2002. The mission consisted of two identical spacecraft following each other in the same orbit, separated by about 220\,km. The orbit characteristics (eccentricity $\approx~0.001$, inclination $\approx 89^{\circ}$) are similar to those from CHAMP. Regarding the satellite altitude, Fig.~\ref{fig:3} illustrates the course of both satellite missions during the investigation period 2003--2015.
Most of the time the GRACE mission was orbiting the Earth at around 490\,km. However, in the last few years of the mission duration (decommission was in November 2017) the altitude decreases more and more rapidly. Comparing the temporal course of the satellite altitude's we can recognize that the two missions are approximately separated by 80--200\,km height difference. Subsequently, the estimated neutral densities ($\rho$) for the lower satellite mission CHAMP must be significantly higher than those from GRACE (cf. Fig.~\ref{fig:7} [c, d]). In either case the density determinations are based on accelerometer measurements ($a$) aboard the satellites, that can be calculated as
\begin{equation}
 \vec{a} = -\frac{1}{2} \rho \left(\sum\limits_{i=1}^{plates} C_{D,i} \frac{A_i}{m} (\vec{v} \cdot \vec{n}_i)\cdot \vec{v} \right).
 \label{eq:rho}
\end{equation}
Therein, $m$ indicates the current satellite mass, $C_{D,i}$ the variable drag coefficient, $A_i$ the effective cross sectional area, $v$ the satellite velocity relative to the co-rotating atmosphere and $n_i$ the unit vector of the individual satellite plate $i$. These satellite specific information were taken from a satellite 9- and 15-plate macro-models for GRACE and CHAMP, respectively~\citep{Bruinsma2003,Bettadpur2007}. For a more detailed description of the neutral density estimation procedure we refer the reader to~\citet{Sutton2008,Doornbos2009,Krauss2013}.
It has to be noted that, due to the ellipticity of the satellite orbits and the long analysis period the neutral densities are normalized to an average altitude of 400\,km ($\rho_{\rm 400}$) and 490\,km ($\rho_{\rm 490}$) for CHAMP and GRACE, respectively. To accomplish this task we used the empirical thermosphere model Jacchia-Bowman 2008~\citep{Bowman2008}.

In~\citet{Krauss2015}, the bias and scale parameters for the proper calibration of the raw GRACE accelerometer measurements were kindly provided by the Centre National d'\`{E}tudes Spatiales (CNES). The obtained calibration parameters were calculated till December 2010. To save energy and to extend the satellite mission the active thermal control system of the GRACE satellites was switched off in April 2011 - resulting in the necessity to revise the processing routine. The method used for this study was elaborated in the course of the joint project SPICE (Environmental space geodesy: detection of changes in glacier mass time-variable gravity) and since then already applied for the latest Earth gravity field release ITSG-Grace2016. It already takes into account, the temperature dependency of the accelerometer (bias and scale factor) and should therefore be suitable to analyze GRACE accelerations during the extended investigation period between 2010 and 2015. For further information about the developed calibration routine the reader is referred to~\citet{Klinger2016} and~\citet{Krauss2018}.

\begin{figure*}[htb]
 \begin{center}
  \includegraphics[width=12cm]{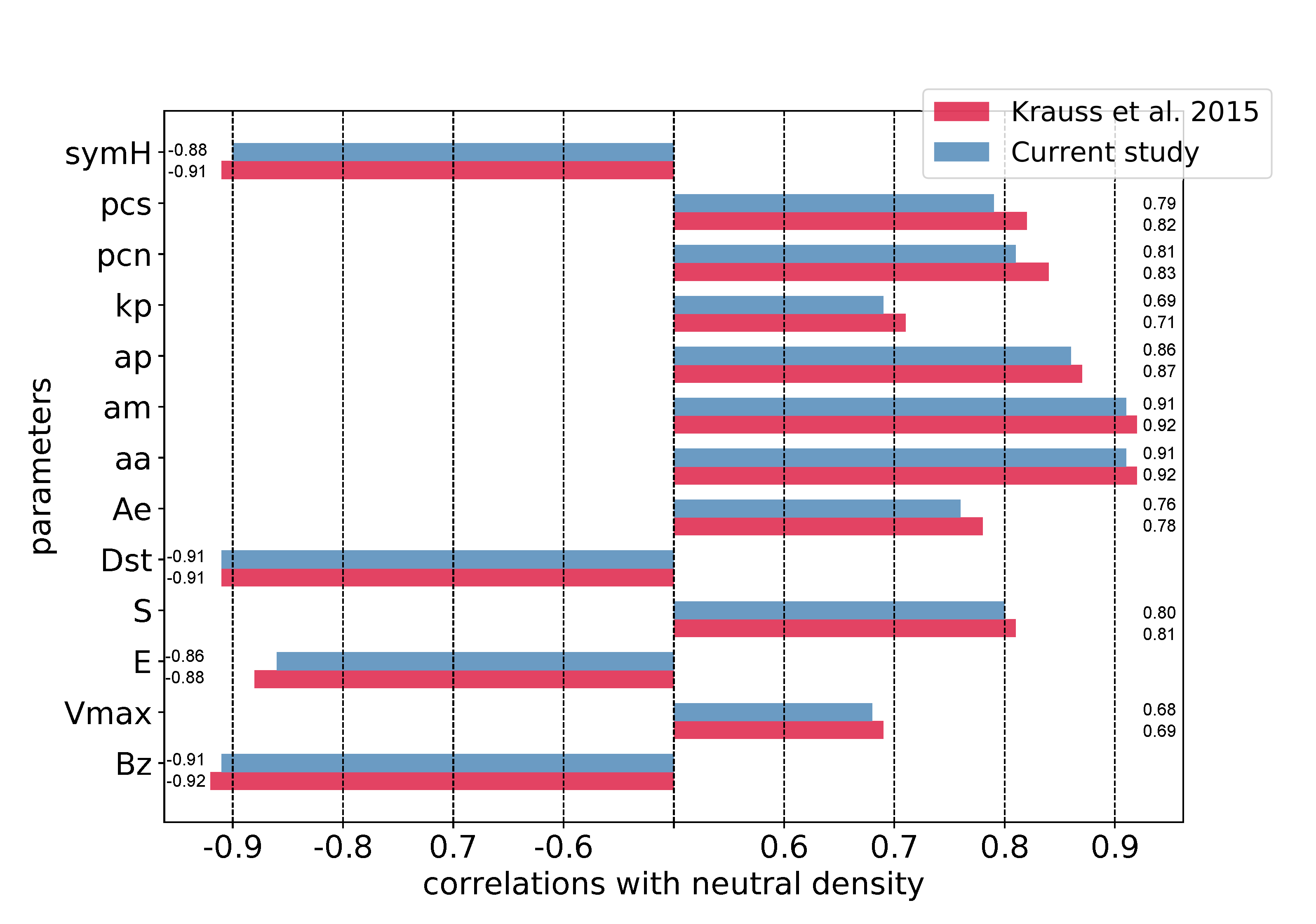}
 \end{center}
\caption{Validation of the new acceleration calibration method [current study] by comparing it with the results from the former~\citep{Krauss2015} study in terms of correlation coefficients with the neutral density. Basis therefore is the former event database from August 2003 to December 2010.}
 \label{fig:4}
\end{figure*}

Since the results of the present study are solely based on the newly formulated calibration method, we initially validated the method by comparing the results with those published in~\citet{Krauss2015}. For this purpose, the data of the former investigation period (July 2003 till December 2010) were reprocessed by using accelerations calibrated with the new routine. Figure~\ref{fig:4} presents the outcome of these computations, in terms of correlation coefficients between the estimated normalized neutral density increase and various analyzed parameters. This includes the magnetic field component $B_{\rm z}$ as well as different geomagnetic indices and corresponding ICME parameters (proxy for the convective electric field [$E$] and for the energy input via the Poynting flux into the magnetosphere [$S$]).
Compared to the predecessor study (red bar), Fig.~\ref{fig:4} reveals that, the correlation coefficients are slightly lower when applying the new calibration method. This can be attributed to the fact, that the new neutral density estimates show a fairly higher noise and scatter behavior. As the differences are within reasonable limits we propose that it is feasible to apply the routine elaborated by \citet{Klinger2016} for the entire period of investigation.

Since~\citet{Krauss2015} has revealed that the correlations between the neutral density enhancement and the $B_{\rm z}$ component of the magnetic field are higher than those with the convective electric field [$E$] or the energy input via the Poynting flux into the magnetosphere [$S$]) we will solely focus on the former one for the further analysis of the in-situ measurements from the ACE spacecraft.
\section{Results}
\subsection{Influence of different solar activity levels on the evaluation results}
\begin{figure*}[htb]
 \begin{center}
  \includegraphics[width=14cm]{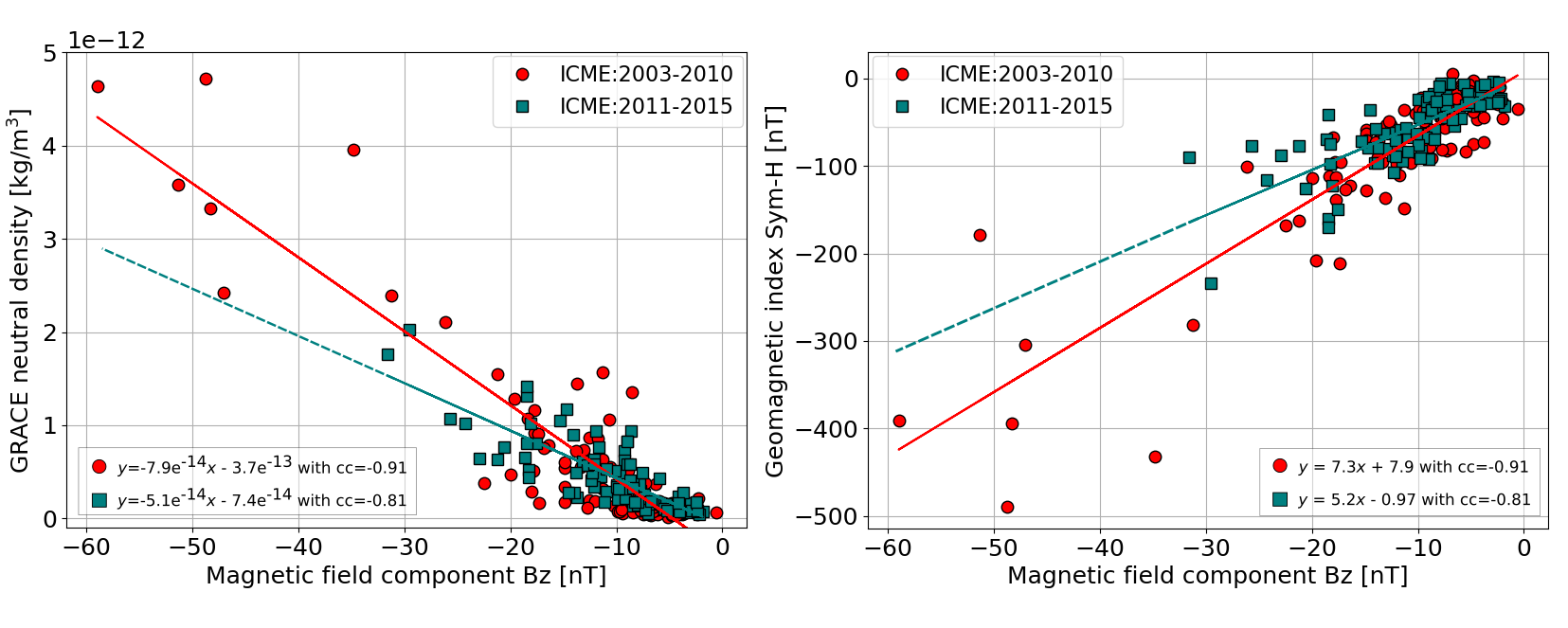}
 \end{center}
\caption{Scatter plots comprising 196 analyzed ICME events. Red circles cover the period of the predecessor study from 2003 till 2010, blue squares the extended analysis period from 2011 to 2015. In the left panel the magnetic field component $B_{\rm z}$ versus GRACE neutral densities are shown; in the right panel $B_{\rm z}$ versus the geomagnetic index Sym-H. For a better visibility, extrapolated regression lines are illustrated in dashed lines, but should be interpreted with care.}
 \label{fig:5}
\end{figure*}

When incorporating the new data from solar cycle 24 we have to keep in mind that, the solar activity levels were clearly different in the time range for 2003--2010 (previous study) and 2011--2015 (new extended period). While 2003--2010 covers the maximum and decaying phase of cycle 23, 2011--2015 comprises the rising and maximum phase of the weak solar cycle 24~\citep{Gopalswamy2015}. Hence, ICME speeds and impacts are expected to be much lower for the extended study than for the results given in~\citet{Krauss2015}. For this reason, we initially carried out a comparison between results obtained for the two separate time periods. Figure~\ref{fig:5} shows scatter plots between the estimated neutral density increase based on GRACE data, the magnetic field component $B_{\rm z}$ measured by ACE as well as the Sym-H index as a representative for the geomagnetic activity level.
For each of the three parameters the expected lower solar activity level during the time period 2011--2015 is apparent. The maximal negative $B_{\rm z}$ magnetic field measured for an ICME event during the extended period is $-31.6$\,nT for an event on October 02, 2013. Due to the absence of extreme ICME events we obtain slightly different results in the regression line trends for the different time ranges. Thus, it has to be taken care when extrapolating data from a limited subset (e.g. time range 2011--2015) of measurements, which might lead to erroneous assumptions (cf. Fig.~\ref{fig:5} dashed lines). However, if we constrain our analysis to smaller CME events (see Fig.~\ref{fig:6}b) we can state that the data of the extended period fits pretty well in the overall picture.\\
In Tab.~\ref{tab:correlation1} an overview of all determined correlation coefficients regarding the increase in the neutral density from GRACE as well as the magnetic field component $B_{\rm z}$ for the complete analysis period is given.
\begin{table}[ht]
 \caption{Correlation coefficient between the various parameters and the increase in neutral density $\Delta \rho_{\rm490}$ (column 2) as well as the $B_{\rm z}$ component measured by ACE (column 3) - time period 2003--2015.}
 \centering
 \begin{tabular}{l c c}
  \hline
  {} & {\bf{$\Delta \rho_{\rm 490}$}} & {\bf{$B_{\rm z}$}} \\
  \hline
  $\Delta \rho_{\rm 490}$  & 1.00 & -0.89 \\
  $B_{\rm z}$  & -0.89 & 1.00 \\
  $v_{\rm max}$  & 0.65 & -0.60 \\
  $E$ & -0.85 & 0.86 \\
  $S$ &  0.78 & 0.77 \\
  Dst  & -0.89 & 0.85 \\
  AE  & 0.71 & -0.73 \\
  a\textsubscript{a}  & 0.89 & -0.87 \\
  a\textsubscript{m}  & 0.89 & -0.88 \\
  a\textsubscript{p}  & 0.84 & -0.83 \\
  k\textsubscript{p}  & 0.67 & -0.75 \\
  PCN  & 0.80 & -0.80 \\
  PCS  & 0.76 & -0.78 \\
  SYM--H & -0.88 &  0.89 \\
  \hline
 \end{tabular}
 \label{tab:correlation1}
\end{table}
The resulting correlations are slightly lower than those obtained for the period 2003--2010 (cf. Fig.~\ref{fig:4}), due to the already mentioned fact that the extended period missed extreme ICME events, which are important in a statistical manner. However, overall the results from both time periods fit together well. Based on that, we safely can extend our statistical analysis from the time period 2003--2010 (cf. Fig.~\ref{fig:4}) to the entire investigation period January 2003 to December 2015.\\
We note that Tab.~\ref{tab:correlation1} only refers to density increases computed for the GRACE satellites. This is because the analyzed CHAMP data lead to similar results due to a nearly 1:1 correlation between GRACE and CHAMP densities (cc=0.97).
\subsection{Investigating the impact of CIR induced geomagnetic storms on the Earth thermosphere}
\begin{figure*}[htb]
 \begin{center}
   \includegraphics[width=14cm]{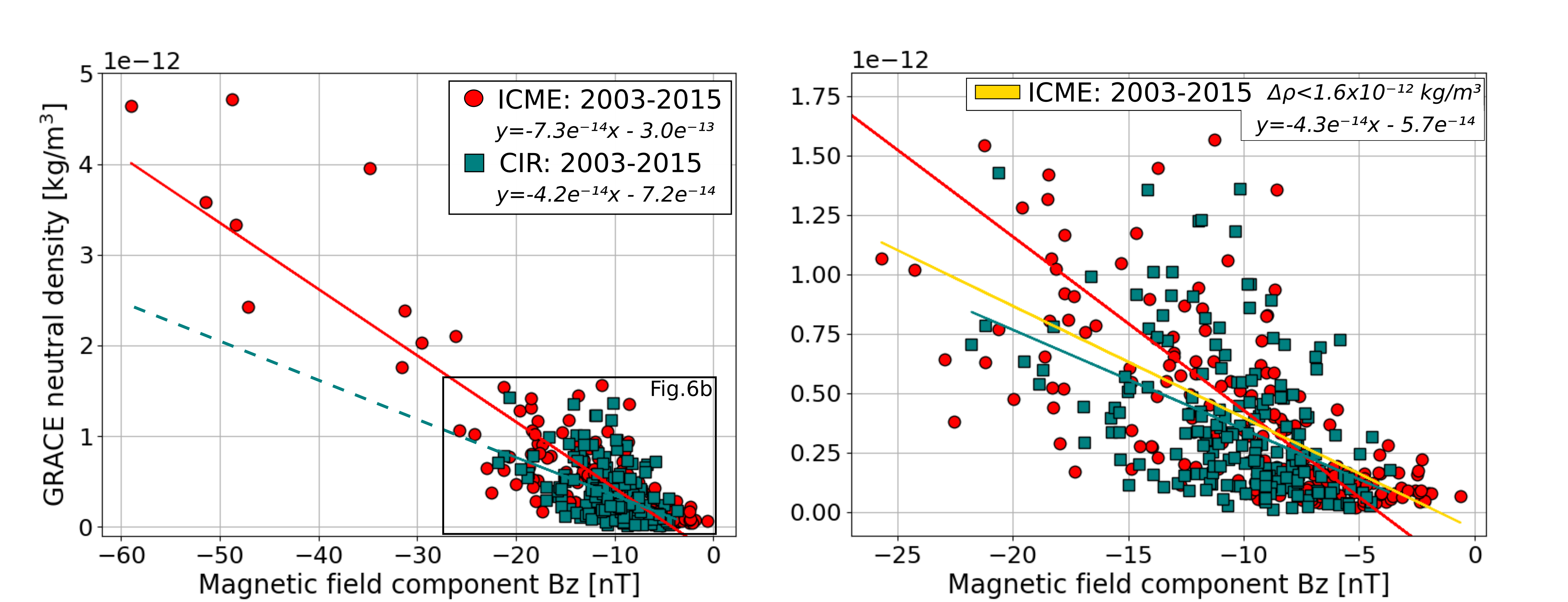}
 \end{center}
\caption{(left to right) scatter plots of the minimum $B_{\rm z}$ component measured by ACE with the increase in neutral density and a magnified version which includes the outcome of the ICME analysis including only ICME events which triggered a density enhancement $\Delta \rho < 1.6\times10^{-12}$~kg\,m$^{-3}$ (yellow line). For a better visibility, extrapolated regression lines are illustrated in dashed lines, but should be interpreted with care.}
 \label{fig:6}
\end{figure*}

In a second step of the study we examined, how CIR induced geomagnetic storms affect the thermospheric density increase. The illustration in Fig.~\ref{fig:6}a represents a scatter plot including all analyzed ICME and CIR events in terms of increasing neutral densities and the maximal negative $B_{\rm z}$ component measured by GRACE and ACE, respectively. CIR induced geomagnetic storms have weaker effects on the thermosphere density than those caused by ICME events. This can be attributed to their overall smaller $B_{\rm z}$ magnetic field as well as to the fact that the relevant energy input into the thermosphere comes from the compression region of the CIR. 
During the analysis period, CIR events cause variations in the magnetic field component $B_{\rm z}$ only up to about -22~{nT} and density increases at GRACE altitudes in the order of approximately $~1.4\times10^{-12}$~kg\,m$^{-3}$. This means that, the influence of CIR events is in the same range of magnitude as the majority of the ICMEs (186 out of 196) and thus can be seen as an extended distribution of the density enhancements related to weak ICMEs. This can be further illustrated by excluding the ten most extreme ICME events, with a density enhancement greater than $~1.6\times10^{-12}$~kg\,m$^{-3}$, in the analysis. This yields a significant better agreement between the ICME and CIR analysis and indicates that the effect of negative $B_{\rm z}$ is not a linear one. The findings are given in Fig.~\ref{fig:6}b, which reveals a magnified view of Fig.~\ref{fig:6}a, and additionally includes the derived regression line from the reduced ICME analysis.\\

CIR events are generally characterized through a turbulent and strong magnetic field due to the compression between the slow and fast solar wind regime. Against this background, the effect on the Earth's thermosphere should agree better with those from ICMEs where the negative $B_{\rm z}$ component is located in the shock-sheath region~(cf.~\citet{Krauss2015}).
\begin{figure*}[htb]
 \begin{center}
   \includegraphics[width=8cm]{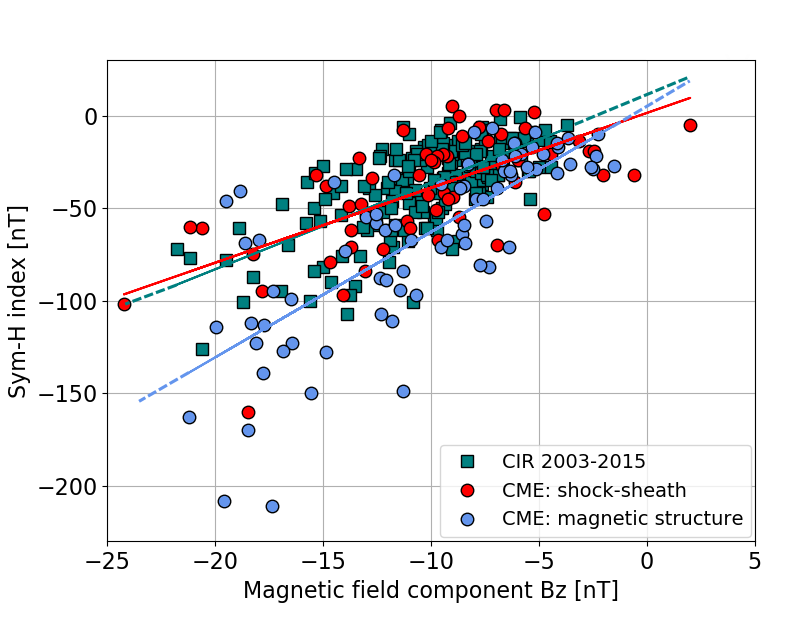}
 \end{center}
\caption{Scatter plot of the minimum $B_{\rm z}$ component measured by ACE with the geomagnetic index Sym-H. Illustrated are events over the entire investigation period 2003--2015: 195 CIR (green squares) and 196 ICME (red/blue circles).}
 \label{fig:7}
\end{figure*}
Figure~\ref{fig:7} shows the relationship between the $B_{\rm z}$ measured by ACE and the geomagnetic activity level on the example of the Sym-H index for all analyzed CIR and ICME events (separated in shock-sheath and magnetic structure). It can be clearly seen that, the impact of CIR induced geomagnetic storms on the Sym-H index is very similar to the effects from shock-sheath regions of an ICME that carry a lower $B_{\rm z}$ than the magnetic structure.
\subsection{Height depended impact of geomagnetic storms induced by ICME and CIR events}
\begin{figure*}[ht]
 \begin{center}
  \includegraphics[width=14cm]{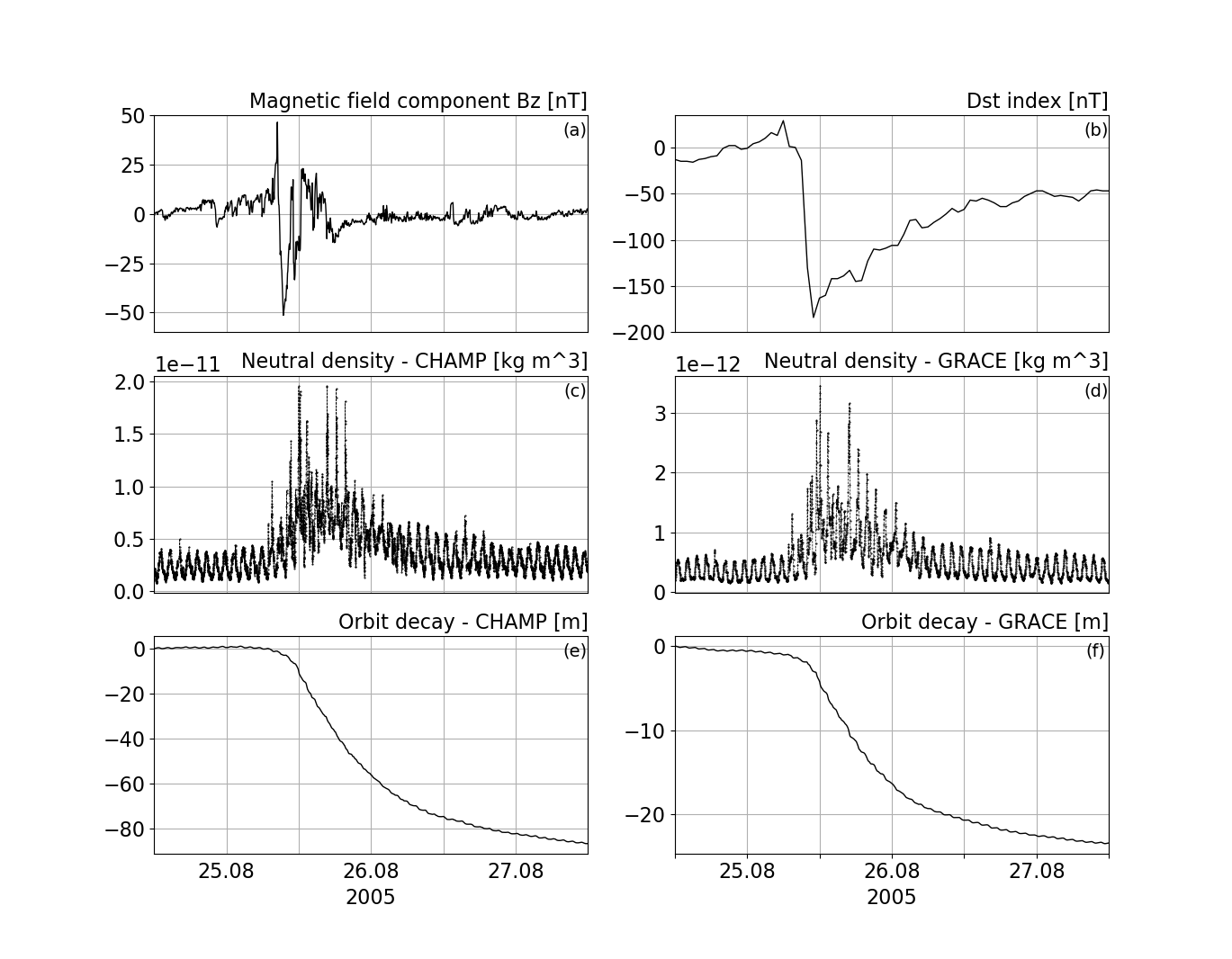}
 \end{center}
 \caption{Overview of various analyzed parameters during an ICME event in August 25, 2005. Including the ACE measurements of the $B_{\rm z}$ component (a), the Dst-index (b), the estimated neutral densities for CHAMP and GRACE (c,d) and finally the determined orbit decay for both satellite missions (e,f).}
 \label{fig:8}
\end{figure*}

The relative density increase, due to ICME and CIR induced geomagnetic storms, is found to be very similar for GRACE or CHAMP. Therefore, all results so far referred solely to the neutral density estimate for the GRACE spacecraft (see Tab.~\ref{tab:correlation1}). However, due to their different orbit heights (cf. Fig.\ref{fig:3}), the satellites are exposed to different absolute atmospheric densities. Examining the subplots Fig.~\ref{fig:8}~[c, d], we see that the estimated densities differ by an order of magnitude. Hence, density increases due to an extreme solar event will have different effects on orbit decays for GRACE or CHAMP.
Following ~\citet{Chen2012} the orbit decay can be calculated in terms of changes in the mean semi--major axis. Figures~\ref{fig:8} summarize various analyzed parameters during an ICME event in August 2005. It can be seen that, due to the large differences of both satellite missions regarding their absolute neutral atmospheric densities, tantamount to their orbital altitude, the orbit decay for GRACE and CHAMP is significantly different, even though it is triggered from the same solar event.
\begin{figure*}[ht]
 \begin{center}
  \includegraphics[width=14cm]{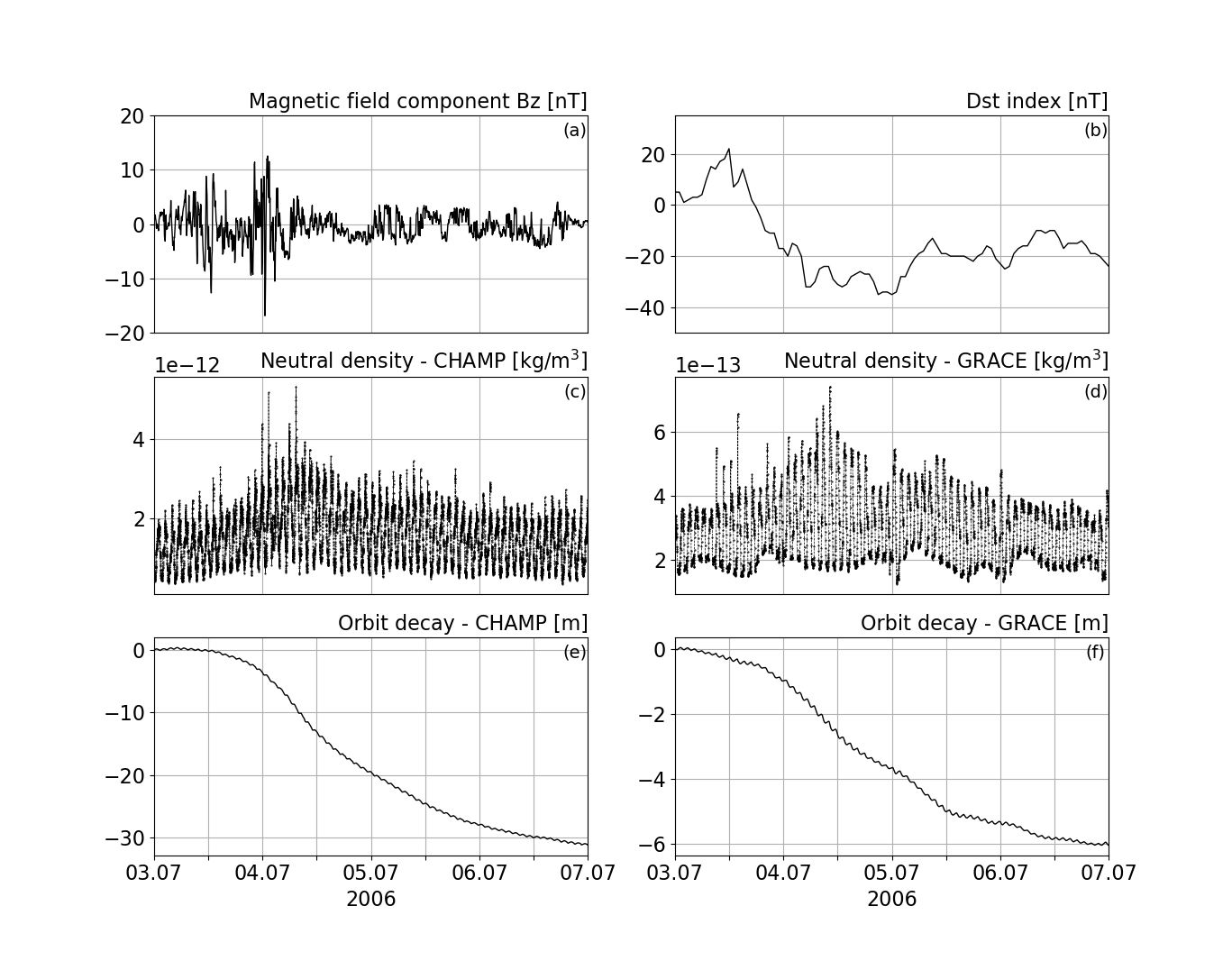}
 \end{center}
 \caption{Overview of the same parameters as shown in Fig.~\ref{fig:8} but for a CIR event in July 03, 2006.}
 \label{fig:9}
\end{figure*}
Regarding the CHAMP orbit decay rates our results confirm the findings from~\citet{Chen2014}. The authors compared the effects of CIR- and CME-induced geomagnetic storms using data from this specific satellite mission and found that both types of solar events might trigger comparable orbit decay rates. Figure~\ref{fig:9} illustrates various analyzed parameters during such an CIR event in July 2006.\\

\begin{figure*}[ht]
 \begin{center}
  \includegraphics[width=10cm]{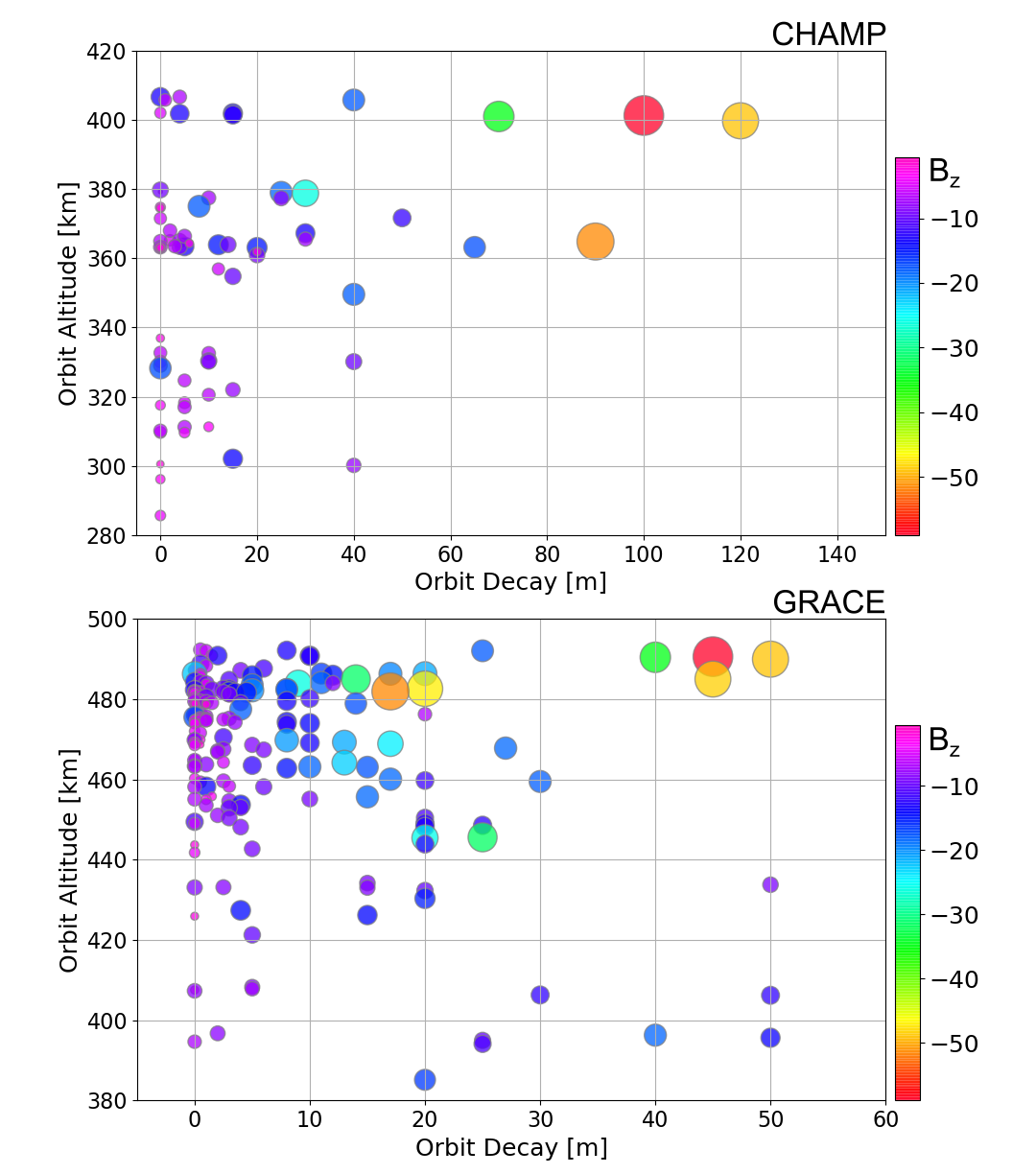}
 \end{center}
 \caption{Orbit decays versus orbital altitude and event strength in terms of $B_{\rm z}$ [nT] measurements for the CHAMP (top) and GRACE spacecraft (bottom). To obtain a better visibility, two average events which triggered large orbit decays due to an extreme low satellite orbital altitude are neglected for the present illustration.}
 \label{fig:10}
\end{figure*}

However, as noted before, the decay rate depends significantly on the actual satellite altitude. For the GRACE orbit, which was located about 150\,km higher than CHAMP, the expected decay is much lower due to the less denser atmosphere. When analyzing the effects of solar events on the orbital altitude of satellites over a 13-years investigation period different changing conditions must be kept in mind: Firstly, due to the varying solar conditions (cf. solar cycle 23, 24), solar events have different impact intensities. Secondly, the actual altitude (cf.~\ref{fig:3}) of a satellite represents a significant factor in the expected orbital decay. Figure~\ref{fig:10} summarizes the impact of nearly 200 ICME events for the CHAMP (upper panel) and the GRACE satellite mission (lower panel) exemplified by the estimated orbital decay, the prevailing satellite altitude and the intensity of the specific solar event in terms of the in-situ measurements of the magnetic field component $B_{\rm z}$ - the larger the bubble the stronger the negative IMF $B_{\rm z}$ event.
Therein, Fig.~\ref{fig:10} reveals that, a large number of ICME events which occurred during the 13-years investigation period had minor effects on the GRACE satellite orbit (around 450--490\,km) but already perceptible implications for the CHAMP mission. Extreme ICME-induced geomagnetic storms may lead to decay rates of 40--50\,m and 90--120\,m for GRACE and CHAMP, respectively. Furthermore, it demonstrates, that during periods of lower orbital altitudes even smaller solar events are sufficient to trigger GRACE orbit decay rates in the order of 20--50\,m. For CHAMP such effects are nearly not existent due to low solar activity at the end of the satellite mission (decommission in 2010). Concerning the time delay between the ICME/CIR arrival and the orbit decay onset an average value of $5 \rm{hrs}\pm 2 \rm{hrs}$ was estimated. 
\clearpage
\section{Discussion}
This paper deals with the analysis of the impact of 196 Earth-directed ICME and 195 CIR events on the Earth upper atmosphere and the direct implications for satellite orbits. The study comprises solar events which occurred in the time period between 2003 and 2015 using in-situ measurements recorded by the ACE spacecraft in the Lagrange point L1 and the two low-Earth orbiting missions CHAMP and GRACE.  As a result, of the additional analyzed CIR events, the extended investigation period (beyond 2010) and the added satellite mission CHAMP, the current study represents an extended analysis of the predecessor study~\citet{Krauss2015}. For the time-based delimitation of the various solar events we rely on the ICME catalog maintained by R\&C and CIR-lists provided by~\citet{Jian2011} and S. Vennerstrom. To process accelerometer measurements beyond 2010 we applied a new calibration routine, whose results were compared with our previous results and calibration method. The analysis has revealed, that the new developed routine delivers reasonable results and is suitable to analyze solar events after the shutdown of the thermal control system in April 2011. Based on these findings we thoroughly examined the effects of solar events between 2003 and 2015 with regard to the occurred thermospheric and geomagnetic response. The derived linear regression parameters confirm our previous findings, which showed a high correlation between the majority geomagnetic indices and ICME parameters with the variations in the atmospheric neutral density. Furthermore, the increase in the neutral densities during an event yield a nearly 1:1 correlation (cc=0.97) between estimates for CHAMP and GRACE satellites.\\
A subsequent investigation on CIR events showed that the impacts on the near-Earth environment are quite similar to those from Earth-directed ICME events - about 90\% of ICMEs are in the same order of magnitude. A direct comparison of ICME and CIR induced geomagnetic storms indicated that both show the same behavior when the main disturbances (minimum in $B_{\rm z}$) are triggered at the time of the shock-sheath region and not in the magnetic structure part.\\
Finally, as proposed in the discussion section of~\citet{Krauss2015}, the current study now addressed an important field of application of space weather, namely the induced satellite orbit decay.
For this purpose, the effects of all ICME and CIR between 2003 and 2015 on the CHAMP and GRACE satellite altitude were estimated. In doing so, we can confirm findings by~\citet{Chen2014} who demonstrated that CIR events have similar effects on the total orbital decay than ICME induced geomagnetic storms. However, when comparing the two satellite missions CHAMP and GRACE it became obvious that the actual orbital height is a crucial factor. One and the same solar events triggers substantial different decay rates (up to 70\,m) at different altitudes.
\acknowledgments
We acknowledge the use of the satellite data from ACE, CHAMP and GRACE as well as the world data center (WDC) for Geomagnetism, Kyoto. ACE data used in this study were obtained through the OMNI database (omniweb.gsfc.nasa.gov). The access to the accelerometer measurements from the CHAMP and GRACE satellites was provided by the Information System and Data Center (ISDC) in Potsdam (isdc.gfz-potsdam.de). The authors additionally acknowledge the use of the ICME list of Richardson and Cane as well as the CIR lists of L.K. Jian and S. Vennerstrom.
Furthermore, we would like to thank B. Klinger (University of Technology Graz) for her support concerning the GRACE accelerometer calibration. Martina Edl contributed to the CIR related data preparation. S.K.\,acknowledges the support by the FFG/ASAP Program under grant no. 847971 (SPICE) and M.T. the FFG/ASAP Program under grant no. 859729 (SWAMI).



\end{document}